\documentclass[12pt]{article}
\usepackage{amssymb}
\usepackage{amsfonts}
\usepackage{amsmath,amsthm}
\usepackage{graphics,epsfig}
\usepackage{subfigure}
\usepackage{graphicx,epsfig, color }
\oddsidemargin 10mm \numberwithin{equation}{section}
\begin{document}
\begin{center}
{{\bf {Canonical quantization of modified non-gauge invariant
Einstein-Maxwell gravity and stability of spherically symmetric
electrostatic stars}}
 \vskip 1 cm { Hossein
Ghaffarnejad\footnote{E-mail address: hghafarnejad@semnan.ac.ir},}} \\
\vskip 0.1 cm {\textit{Faculty of Physics, Semnan University, P.C.
35131-19111, Semnan, Iran}
 } \vskip 0.1 cm
\end{center}
\begin{abstract}
We consider a non-minimally coupled Einstein-Maxwell gravity with
no $U(1)$ symmetry property to study stability of an electrostatic
star via canonical quantization approach and obtain that the
stability is free of gauge field effects. By calculating the
Hamiltonian density of the stellar system we show that the
corresponding Wheeler-DeWitt wave functional is similar to a
simple harmonic quantum Oscillator for which a non zero ADM mass
of the system causes a quantization condition on the metric
fields. Probability wave packets are described by the Hermit
polynomials. Our mathematical calculations show that in this
approach of quantum gravity the metric fields are regular for all
values of the electric potential and so the quantized spacetime
has not both of event and apparent horizons. The most probability
of the quantized line element is for ground state of the system.
To check validation of the model we use Bohr`s correspondence
principal and generate directly semi classical approach of the
quantized metric states at large quantum numbers where they reach
to Schwarzschild like metric according to the Birkhoff's theorem.
Also we check that the generated semi classical solutions are
satisfied exact classical metric solutions which are obtained from
Euler Lagrange equations. We show that `charge to mass ratio` of
the electrostatic star is a constant defined by the coupling
constant of the model and it is in accord to other alternative
approaches.
\end{abstract}
\section{Introduction}
The famous singularity theorems show that the classical theory of
general relativity is incomplete, because these are causal and so
under general conditions are unavoidable (see Chapter 9 in ref
\cite{wald}). In fact these singularities reach to some infinite
values for energy density and so curvature of curved spacetimes in
a gravitational system under consideration. Conceptually,
physicists hope the singularity problems  resolved via
generalization of the quantum field theory to  gravity or
equivalently to curved spacetimes. This idea comes from history of
the classical mechanics which happened at first of twentieth
century by extending it to the quantum mechanics and quantum field
theory in flat Minkowski spacetime respectively for
nonrelativistic and relativistic particles. In fact, this is the
main reason why we are looking for full quantum gravity.
 In general, to construct full
quantum theory of gravity, there are several approaches which are
formulated and basically are different from each others. In short
they are as follows \cite{Bir}: (a) Quantum general relativity
which is application of `Dirac`s quantization rules` to classical
general relativity. These approaches are done with two different
methods called as `covariant approaches` which means utilizes of
four-dimensional covariance for studying of the system under
consideration, such that perturbation theory \cite{Devpar},
\cite{Birr} or path integral methods \cite{Fior} and `canonical
approaches` where one makes a Hamiltonian formalism and produces
some appropriate canonical variables with corresponding conjugate
momenta. Some examples to the latter formalism are quantum
geometrodynamics \cite{Dewitt} (see also \cite{Pau} for
applications in quantum cosmology with supersymmetric perspective)
and loop quantum gravity \cite{Rovi}. (b) String/M-theory which is
main approach to construct a unifying quantum framework of all
interactions of gravitational and matter fields (see for instance
\cite{Bar} and \cite{Mau}). (c) Quantum gravity from topological
quantum field
theory approaches (see for instance \cite{John} and \cite{Lau}).\\
Interaction of quantum matter fields with black holes which
produces the well known Hawking radiation \cite{Haw1}, \cite{Haw2}
which means evaporation of the black holes, predicts that the full
quantum gravity is dominated at scales below the Planck scales
$\ell_p\approx1.62\times10^{-33}cm$ (see for instance Page 2 of
ref. \cite{Bir}) and for scales larger than of $\ell_p$ the
geometry/gravity behaves classically but not other matter fields.
In the latter case the matter fields behave as quantum fields
which renormalized expectation values of whose stress tensor
operator are used as right side of the Einstein`s metric equation
$G_{\mu\nu}=-8\pi G<\hat{T}_{\mu\nu}>_{ren}.$ This equation is
called backreaction equation of the matter quantum fields
effecting on the classical metric field and its solutions give us
modified background classical metric (see Chapter 14 in
\cite{wald}). This approach of quantum gravity is called `quantum
field theories in curved space (see \cite{Devpar} and
\cite{Birr}). Many authors published several papers by
applications the above mentioned three different approaches of
quantum gravity and due to the wide range of published works, we
only mention a few examples of them here. Gerard`t Hooft explained
in the work \cite{Ger} that to address the problem of information
conservation, the usual expression for the temperature of
Hawking's radiation is off by a factor 2. Authors of the work
\cite{Ham} showed that how can black hole information recovered
from gravitational waves via correlations with the Hawking
radiation. Rubio et al showed in the paper \cite{Rub} conditions
for stability of a regular black hole in presence of Hawking
radiation. As an application of loop quantum gravity author`s of
the work \cite{kim} showed that string T-duality effects avoid
singularity of dust collapsing matter with Hawking radiations in
absence of its backreaction stress tensor expectation values
corrections. Stability of quantum Schwarzschild-de Sitter black
hole in presence of back reaction stress tensor of Hawking
radiation was studied previously by the author in ref.
\cite{Ghaf1} where evaporated quantum Schwarzschild de Sitter
black hole reaches into stable remnant mini black hole. As
application of canonical quantum gravity approach Neto et al
calculated tunneling probability in ref \cite{Net} for birth of
FLRW universe in flat and closed form in presence of radiation and
useful geometric potentials by solving the Wheeler DeWitt wave
equation. As an application of loop quantum gravity approach,
authors of the work \cite{Mot} are studied in effects of loop
quantum geometry on birth of universe via Vilenkin's tunneling
wavefunction proposal.  In the framework of the Wheeler-DeWitt
theory of quantum gravity author of the work \cite{She} find birth
of the universe via metric signature transition. Author
investigated previously,  quantum cosmology of flat
Robertson-Walker spacetime in ref \cite{12} for a modified
scalar-vector-tensor Brans-Dicke gravity by solving the
Wheeler-DeWitt wave equation. In that work it was shown that the
Wheeler DeWitt eigenfunctions are described via the two
dimensional quantum harmonic Oscillator wave packets and so there
is not naked singularity which appears in the classical cosmology
at birth of begin of the RW universe. In this work we like
investigate quantum stability of spherically symmetric time
independent spacetime supported by a modified Einstein-Maxwell
gravity in the canonical quantum gravity approach as follows.
Content of this paper is as
follows:\\
In section 2 we describe shortly, generalized non-minimally
coupled Einstein-Maxwell gravity. Then we obtain Lagrangian and
Hamiltonian densities of the model for a general spherically
symmetric static metric in a Schwarzschild frame. By applying
Dirac`s canonical quantization operators for canonical momentum of
the fields, we find Wheeler-Dewitt wave equation of the system and
solve it, in the section 3. We show that Wheeler-DeWitt wave
solution of the system is described by simple harmonic quantum
Oscillator wave packets. Energy eigenvalues of the system reach to
a quantization condition of the background metric for non-zero ADM
mass of the system. In the section 4 we use Bohr`s correspondence
principal  to show that our perfect quantum gravity proposal for a
spherically symmetric static curved space time with no vanishing
ADM mass reaches to a Schwarzschild-like metric at large quantum
numbers covering the Birkhoff`s theorem. In this section we
calculated net electric charge of the model such that it depends
to the ADM mass of the system linearly.  In section 5 we
calculated exact analytic metric solutions by solving the Euler
Lagrange equations of the fields. We show in two dimensional phase
space that metric field is regular for all values of the electric
potential of the system. This leads us clime that the metric
solutions describe a electrostatic star not a black hole. Section
6 dedicated to concluding remarks and outlook of the work.
\section{The gravity model }
Let us first say about motivation and importance of the following
exotic Einstein-Maxwell gravity model which why we will use in
this work? It is not a secret to anyone that magnetic fields
penetrate throughout the universe and play an important role in
multitude of astrophysical situations. The magnetic field of our
galaxy plays an important role in the dynamics of the galaxy and
making stars , pulsars, and black hole and other astrophysical
objects. To know what is the origin and importance of cosmic
magnetic fields one can see \cite{park}, \cite{Zel} and
\cite{Ell}. Many astrophysicist believe that the galactic magnetic
field are generated and maintained by dynamo action (see for
instance\cite{park2}). The dynamo mechanism is an amplification
mechanism and requires a seed magnetic field which may be coming
from the rotation of spiral galaxies and so on. One can see
\cite{Vil1}, \cite{Vil2} and \cite{Hog} and references therein to
study other more exotic scenarios for origin of the cosmic
magnetic fields.  However at the present situations there is no
compelling  unique mechanism has yet been suggested for the origin
of primeval magnetic fields. There are three reasons that why we
believe that the cosmic inflation is a prime candidate for the
production of primordial magnetic fields \cite{MS}: (a) Inflation
provides the kinematic means of producing very long wavelength
effects at very early times through microphysical processes
operating on scales less that the Hubble radius. (b) Inflation
provides the dynamical means of exiting these long wavelength
electromagnetic waves and (c) During inflation the Universe is
devoid of charged plasma and is not good conductor, so that
magnetic flux is not necessarily conserved. In this view it seems
that for description of theoretical feature of the primeval
magnetic fields which are an initial source of the cosmic magnetic
field, one can use well known minimally coupling Einstein Maxwell
gravity such that
\begin{equation}I=\frac{1}{16\pi G}\int dx^4\sqrt{g}\bigg(R-\frac{1}{4}F _{\mu\nu}F^{\mu\nu}\bigg).\end{equation}
  But it is well known that this
model is a pure $U(1)$ gauge invariance theory and so the
electromagnetic field lagrangian density is conformally invariant.
For such a model the magnetic field always decreases with inverse
of square of scale factor of an expanding FRW cosmology,
regardless of plasma effects \cite{MS}. This causes to suppress
the corresponding energy density so that in the de Sitter phase
inflation the vacuum energy density is just dominated. Thus if we
want to be dominated the magnetic energy density at duration of
inflation and perhaps after the reheating phase the conformal
invariance of electromagnetism must be broken to produce
appreciable primeval magnetic flux. We should remember that nature
shows no sign of being conformally invariant and so we are free to
propose other alternatives instead of the above mentioned Einstein
Maxwell theory. In ref. \cite{MS} several alternatives are
proposed and so we use one of them here such that
\begin{equation}\label{action}I=-\int
dx^4\sqrt{g}\bigg[\frac{1}{4}F_{\mu\nu}F^{\mu\nu}+\frac{\alpha}{2}A^2R+\frac{\beta}{2}R_{\mu\nu}A^\mu
A^\nu\bigg],
\end{equation}where $g$ is absolute value of determinant of the metric field and anti symmetric electromagnetic tensor
field $F_{\mu\nu}$ is defined versus the partial derivatives of
the four vector electromagnetic potential $A_{\mu}$ as follows.
\begin{equation} \label{fmunu}
F_{\mu\nu}=\nabla_\mu A_\nu-\nabla_\nu A_\mu=\partial_\mu
A_\nu-\partial_\nu A_\mu
\end{equation} with $A^2=g_{\mu\nu}A^\mu A^\nu$ and $R_{\mu\nu}$ is Ricci
tensor. It is easy to check that this model is not gauge invariant
same as one which is given in \cite{GHN} where the action
functional remain unchanged by transforming $A_{\mu}\to
A_{\mu}+\gamma\partial_\mu \xi(x^\alpha)$. In this transformation
$\xi$ is gauge field with charge $\gamma$ for which $F_{\mu\nu}\to
F_{\mu\nu}$ namely remains unchanged. One can show that the action
functional (\ref{action}) is transformed to the following form
under the gauge transformation $A_\mu\to A_\mu+\gamma\partial_\mu
\xi(x^\alpha).$
\begin{equation}\label{lineg}I(\gamma)=-\int
dx^4\sqrt{g}\bigg[\frac{1}{4}F_{\mu\nu}F^{\mu\nu}+\frac{\alpha}{2}A^2R+\frac{\beta}{2}R_{\mu\nu}A^\mu
A^\nu\bigg]\end{equation}$$+\alpha\gamma\int
dx^4\sqrt{g}R(A^\mu\partial_\mu\xi+\frac{\gamma}{2}g^{\mu\nu}\partial_\mu\xi\partial_\nu\xi)$$$$+\beta\gamma
\int dx^4\sqrt{g}(R_{\mu\nu}A^\mu
\partial^\nu\xi+\frac{\gamma}{2}R_{\mu\nu}\partial^\mu\xi\partial^\nu\xi)$$
 where the canonical momentum of the gauge field defined by
 \begin{equation}\Pi^\mu_\xi=\frac{\partial L}{\partial(\partial_\mu\xi)}=
 \alpha\gamma\sqrt{g}RA^\mu+\alpha\gamma^2\partial^\mu\xi+\beta\gamma R^{\mu\nu}+\beta\gamma^2 R^{\mu\nu}\partial_\nu\xi
 \end{equation} is a constant of the system because the action functional is free of $\xi.$
  This property gives us an opportunity in study of canonical quantization of
 a spherically symmetric time-independent static metric
\begin{equation}\label{line}ds^2=-X(r)dt^2+Y(r)dr^2+r^2(d\theta^2+\sin^2\theta d\varphi^2)\end{equation}
where we obtain suitable conditions on its quantum stability. In
the static form of the system the vector potential has just time
component $A_t(r)=\phi(r).$ To investigate canonical quantum
gravity of the line element (\ref{line}) we use a Schwarzschild
frame for simplicity where
\begin{equation}\label{sch}Y(r)=\frac{1}{X(r)}\end{equation}  and substitute $A_t(r)=\phi(r)$ and line element (\ref{line})
into the action functional (\ref{lineg}) for which the
corresponding Lagrangian density reads
\begin{equation}\label{lag}
L=2\pi\bigg\{\dot{\phi}^2+
\frac{\beta}{2}\phi^2\bigg[-\frac{2\dot{X}}{X}+\frac{1}{X}-
1\bigg]-\frac{\phi\Pi^t_\xi(\beta+\gamma)}{\beta\gamma}\bigg\}\end{equation}
in which
\begin{equation}\label{Pi}\Pi_\xi^t=\beta\gamma\bigg(1-\frac{1}{X}\bigg)(\phi+\gamma\partial_t\xi)=constant.\end{equation}
 Here we defined $\dot{~}$ as
logarithmic derivative of the radial coordinate as
\begin{equation}\label{dot}\dot{~}=\frac{d}{d\tau}=r\frac{d}{dr}=\frac{d}{d\ln(r/D)}\end{equation}
in which $D$ is a suitable length parameter. Also we use the
ansatz \begin{equation}\alpha=-\frac{\beta}{2}\end{equation} to
remove second order derivatives of the $X$ field coming from the
Ricci scalar and Ricci tensor.
 By defining the canonical momentum of the fields
$X$ and $\phi$ such that
\begin{equation}\label{27}\Pi_{\phi}=\frac{\partial L}{\partial \dot{\phi}}=4\pi\dot{\phi},~~~\Pi_{X}=\frac{\partial L}{\partial
\dot{X}}=-\frac{2\pi\beta\phi^2}{X}\end{equation}and using the
definition of the Hamiltonian density of the fields
\begin{equation}H=\Pi_{\phi}\dot{\phi}+\Pi_{X}\dot{X}+\Pi_\xi^t\partial_t\xi-L\end{equation}
one can show that
\begin{equation}H=\frac{\Pi_\phi^2}{8\pi}+\pi\beta\phi^2
\bigg(\frac{X-1}{X}\bigg)+\frac{X}{(X-1)}\frac{(\Pi^t_\xi)^2}{\beta\gamma^2}
+2\pi\phi\Pi_\xi^t\bigg(\frac{1}{\beta}+\frac{1}{\gamma}-\frac{1}{2\pi}\bigg)\end{equation}
 where the metric field $X$ behaves as parameter because its momentum conjugate $\Pi_X$ is not appeared in the
 Hamiltonian density.
 By defining dimensionless mass $m$ and locally space dependent frequency $\omega(X)$ as
\begin{equation}\label{fer}
m=4\pi,~~~\omega(X)=\sqrt{\frac{\beta}{2}\bigg(1-\frac{1}{X}\bigg)}\end{equation}
  we can rewrite the above Hamiltonian density such that
\begin{equation}\bar{H}=\frac{\Pi_{\phi}^2}{2m}+\frac{1}{2}m\omega^2\phi^2+\frac{(\Pi_\xi^t)^2}{2\omega^2\gamma^2}
+2\pi\phi\Pi_\xi^t\bigg(\frac{1}{\beta}+\frac{1}{\gamma}-\frac{1}{2\pi}\bigg)
\end{equation} where the first term is same as kinetic energy density of a free particle and second term is same as harmonic Oscillator
energy potential density. Two other last terms are interaction
part between the electromagnetic field $\phi$ and the gauge field
$\xi.$ In the canonical quantization approach the last term with
linear dependent momentum of the fields is called momentum
constraint condition
\begin{equation}\label{H1}H_1=2\pi\bigg(\frac{1}{\beta}+\frac{1}{\gamma}-\frac{1}{2\pi}\bigg)\phi\Pi_\xi^t\end{equation}and
all the other second orders of momentum of the fields is called as
Hamiltonian constraint
\begin{equation}\label{H0}H_0=\frac{\Pi_{\phi}^2}{2m}+\frac{1}{2}m\omega^2\phi^2+\frac{(\Pi_\xi^t)^2}{2\omega^2\gamma^2}.\end{equation}
By looking at the frequency equation (\ref{fer}) and the gauge
field momentum (\ref{Pi}) one can infer that
\begin{equation}\lim_{X\to1}\Pi_\xi^t=0,~~~\lim_{X\to1}\omega(X)=0\end{equation} which means that the above Hamiltonian
constraint reduces for free particles with mass  $m=4\pi$ at flat
regions $X\to1$ of the spacetime while in the curved region of
spacetime $X\nrightarrow1$ they are not vanish and so the system
of the fields in minisuperspace $(\phi,\xi,X)$ reaches to a simple
Harmonic Oscillator with non-vanishing frequency $\omega(X)\neq0$
and $\Pi_\xi^t\neq0.$ Furthermore we know that in the classical
regimes of the metric field for spherically symmetric static
spacetime the horizon position is determined by solving the null
hypersurface equation
$g^{rr}\partial_r\Sigma(r)\partial_r\Sigma(r)=0.$ For the line
element (\ref{line}) in the Schwarzschild frame (\ref{sch}) this
equation reads to particular hypersurface $X\to0$ for which with
$\beta<0$ we have $\omega\to\infty.$  In the following section we
use Dirac canonical quantization rules and investigate quantum
stability conditions of the spacetime on the spacial constant
hypersurfaces $X=constant$. We are permit to define a metric
dependent frequency defined by (\ref{fer}) because in the above
Hamiltonian density the momentum conjugate of the metric field $X$
is not appeared in the Hamiltonian density and so in the Wheeler
DeWitt wave equation given in the subsequent section the metric
field $X$ should plays as a parameter in minisuperspace models. In
other words this means that the frequency given by (\ref{fer}) is
not changed on the minisuperspace $X=constant.$ In the following
section we apply to quantize this hamiltonian and generate
eigenvalues and eigenfunctions on a particular minisuperspace
$X=constant.$
\section{Quantization of spacetime}
 In the canonical quantization approach, Dirac`s quantization rules for the canonical momentum operators of the
 fields read \begin{equation}\label{31}\hat{X}=i\frac{\delta}{\delta\Pi_X},~~~\hat{\Pi}_{X}=-i\frac{\delta}{\delta
 X},~~~\hat{\phi}=i\frac{\delta}{\delta\Pi_{\phi}},~~~
 \hat{\Pi}_{\phi}=-i\frac{\delta}{\delta\phi},~~\end{equation} and
 \begin{equation}\hat{\xi}=i\frac{\delta}{\delta\Pi_\xi^t},~~~\hat{\Pi_\xi^t}=-i\frac{\delta}{\delta \xi}\end{equation}
with units $\hbar=G=1$ for which the Wheeler-DeWitt wave solution
$\Psi(X,\phi,\xi)$ should obey  the Hamiltonian and momentum
constrained conditions synchronously such that
\begin{equation}\label{H01}\hat{H}_0\Psi=M_{ADM}\Psi,~~~\hat{H}_1\Psi=0\end{equation}  where we assumed the ADM mass $M_{ADM}$
of the curved spacetime (\ref{line}) to be non-zero.
 By substituting the operators
(\ref{31}) into the constraint conditions (\ref{H1}) and
(\ref{H0}) then the Wheeler DeWitt equation called with
(\ref{H01}) reads to the following form in the minisuperspace
$(\phi,\xi,X).$
\begin{equation}\label{wdw}\frac{\delta}{\delta\xi}\Psi(X,\phi,\xi)=0,~~
~~\bigg[\frac{\delta^2}{\delta\phi^2}-m^2\omega^2\phi^2+
\frac{m}{\gamma^2\omega^2}\frac{\delta^2}{\delta\xi^2}+2mM_{ADM}\bigg]\Psi(X,\phi,\xi)=0\end{equation}
where the left side equation reads that the Wheeler DeWitt wave
should be independent of the gauge field $\xi$ and so we
substitute $\Psi(\phi,X)$ into the right side equation such that
\begin{equation}\label{H010}\bigg[\frac{\delta^2}{\delta\phi^2}-m^2\omega^2\phi^2+2mM_{ADM}\bigg]\Psi(X,\phi)=0\end{equation}
in which $X$ is a constant parametric field coming from the
frequency $\omega(X).$ This means that the above equation is one
dimensional second order linear differential equation for $\phi$.
By setting \begin{equation}\label{def}
y=\frac{\phi}{\phi_0},~~~\phi_0^2=\frac{1}{m\omega},~~~
~\epsilon=\frac{2M_{ADM}}{\omega},~~~P(y)=\Psi(X,\phi)\end{equation}
the equation (\ref{H010}) reads
\begin{equation}\label{36}\bigg(\frac{d^2}{dy^2}-y^2+\epsilon\bigg)P(y)=0.\end{equation}
  Using the method given by the
ref. \cite{Gasi}, one can show that the equation (\ref{36})
reduces  to the well known Hermit equation
\begin{equation}\frac{d^2h(y)}{dy^2}-2y\frac{dh(y)}{dy}+(\epsilon-1)h(y)=0\end{equation} with \begin{equation}
P(y)=e^{-y^2/2}h(y)\end{equation} which with quantization
condition
\begin{equation}\epsilon-1=2n,~~~n=0,1,2,3,\cdots\end{equation} the Hermit solutions $h(r)$ reduce to the Hermit polynomials
\begin{equation}h_n(y)=(-1)^ne^{y^2}\frac{d^n}{dy^n}e^{-y^2}.\end{equation} By regarding the normalization condition on the Hermit polynomials
at last we can write the eigenfunctions of the quantum Harmonic
Oscillator as follows.
\begin{equation}P_n(y)=C_n\exp(-y^2/2)h_n(y),~~~C_n=\frac{1}{\sqrt{2^nn!\sqrt{\pi}}}
\end{equation} in which $C_n$ is normalization coefficient.
 By substituting the quantization
condition $\epsilon=2n+1$ into the definition $\epsilon$ given by
(\ref{def}) we obtain eigenenergies of the system as
\begin{equation}\label{MADM}M_{ADM}=(n+\frac{1}{2})\omega\end{equation} in which $\omega$
should be substituted from (\ref{fer}). If the frequency $\omega$
is known on a fixed minisuperspace $X$, then this equation gives
quantization condition of the ADM mass or energy content of the
spacetime. On the other hand, by assuming that the total ADM mass
or energy of the spacetime is known, then this equation gives
again quantization condition on the background metric of spacetime
$X$. In the latter case it is enough we replace equation of
$\omega(X)$ given by (\ref{fer}) into (\ref{MADM}) to reproduce
\begin{equation}\label{Xn}X_n=\bigg(1-\frac{8M^2_{ADM}}{\beta(2n+1)^2}\bigg)^{-1},~~~Y_n=X_n^{-1}=\bigg(1-\frac{8M^2_{ADM}}{\beta(2n+1)^2}\bigg)
.\end{equation} This remembers  us  the equivalence principle
accepted in general relativity that matter creates geometry and
vice versa. Now we check quantization condition of the event
horizon and apparent horizon which possibly maybe to being. In the
classical regimes of the metric fields we know that for static
spacetime with line element (\ref{line}), position of the event
horizon and the apparent  horizon is obtained by solving the
equations $g_{tt}=X=0$ and $g^{rr}=\frac{1}{Y}=X=0$ respectively.
However it is easy to see that in the quantum version of the
spacetime the equation $X_n=0$ given by (\ref{Xn}) has not any
solutions for a particular $n.$ This means that in the Wheeler
DeWitt canonical quantum gravity approach singularity of a
spherically symmetric spacetime resolves and the metric field is
regular throughout the spacetime. Source of this quantization
condition comes from the interacting part of the action functional
(\ref{action}) for particular case $\beta=-2\alpha$ where electric
potential $\phi$ plays role of position of a geometric particle
which behaves as quantum harmonic Oscillator with rest mass
$m=4\pi$. Thus one can infer that one of motivations of the
presented gravity model can become this result: namely removing of
black hole metric singularity in the static regime so that the
metric define line element of a electrostatic star. In this
minisuperspace quantum gravity approach which is free of
coordinate systems, we do not encounter the causal singularities
of spacetime that we encounter in studying the geometry of
spacetime at the classical level and by solving Einstein's
equations. This is other motivation for the model under
consideration in the canonical quantum gravity approach.  Other
argument which we can be obtained from our calculations is this:
Although the model under consideration is not gauge invariant at
all but in the canonical quantum gravity approach we obtained that
the quantization condition of the spacetime is free of effects of
the gauge field. In other words the Wheeler DeWitt probability
wave of the spacetime is not dependent to value of gauge field and
so our results are correct and valid for every arbitrary used
gauge field.
  We end this  section by giving uncertainty relation on the electric potential field such that \begin{equation}
\delta\phi\delta\Pi_{\phi}=\bigg(n+\frac{1}{2}\bigg)\end{equation}
in which
\begin{equation}\delta\phi=\sqrt{<\phi^2>-<\phi>^2}\end{equation} and
 \begin{equation}\delta\Pi_\phi=\sqrt{<\Pi_\phi^2>-<\Pi_\phi>^2}\end{equation}
are obtained by calculating expectation values of the quantities
$\langle\hat{\phi}^2\rangle\neq0$ and $\langle\hat{\phi}\rangle=0$
with the corresponding momentum expectation values
$\langle\hat{\Pi}_\phi^2\rangle\neq0$ and
$\langle\hat{\Pi}_\phi\rangle=0.$ This is done similar to
calculations which we use in the ordinary quantum mechanics
calculations. In the next we answer to important question such
that can we do generate semi classical metric solutions directly
from the above pure quantization calculations?
\section{Correspondence principle }
It is well known that the Niels Bohr presented at a first time in
1920, the `correspondence principal` to disambiguate of some
possible doubts in the old ordinary quantum mechanics which was
from point of view by some scientist. In this approach the
correspondence principle states that the behavior of systems
described by the quantum theory reproduces classical physics in
the limit of large quantum numbers. The term codifies the idea
that a new theory should reproduce under some conditions the
results of older well-established theories in those domains where
the old theories work. This concept is somewhat different from the
requirement of a formal limit under which the new theory reduces
to the older, thanks to the existence of a deformation parameter.
It was by following this principle that the Einstein's special
relativity was shown that reaches to the Galilean relativity at
low speeds and the general theory of relativity also leads to the
Newton's theory of gravity. We are now try to generate classical
approach of our obtained quantum solutions (\ref{Xn}) in weak
field limits $(n\to\infty)$ by regarding the correspondence
principle. One can infer that for large quantum numbers $n>>1$ the
quantized metric field $X_n$ given by (\ref{Xn}) approaches to the
following approximation
\begin{equation}\lim_{n>>1} X_n\sim1+\frac{8M^2}{\beta(2n+1)^2}\end{equation}
which for $\beta<0$ can be compared with the Schwarzschild form
$1-\frac{2MG}{c^2 r}$ if we set
\begin{equation}\label{rn}r_n=r_0\big(n+\frac{1}{2}\big)^2,~~~r_0=-\frac{\beta G}{Mc^2},~~~\beta<0.\end{equation}
where we suppress the $ADM$ subscript for simplicity. For large
quantum number $n$ the equation (\ref{rn}) reduces to a continuous
quantity $r_n\sim r_0n^2$ as Schwarzschild radius because.
\begin{equation}\lim_{n\to\infty} \frac{\delta
r_n}{r_n}=\frac{2}{n}\sim0,~~~\delta r_n=r_{n+1}-r_n.
\end{equation}  This infers us that the model in weak field limits confirms the `Birkhoff's theorem` in general relativity which states
that any spherically symmetric solution of the vacuum field
equations must be static and asymptotically flat. This means that
the exterior solution (i.e. the spacetime outside of a spherical,
non-rotating, gravitating body) must be given by the Schwarzschild
metric. The argument above for $n>>1$ is in fact one of important
implications of the model which satisfies the correspondence
principle. Other implications of the model is that the model
covers the `cosmic censorship hypothesis` such that there is not a
causal singularity $r=0$ in the non-perturbation approach given in
the previous section while in the classical general relativity
approach there must be a closed surface (the horizon) which covers
casual singularity. As seen above we chose $\beta<0$ to generate
Schwarzschild radius of the line element (\ref{line}) in the weak
field approach of the model but we should note that $\beta>0$ is
also analytic continuation of the $\beta$ parameter for which
$r_0<0$ is still a physical quantity. To show that how this done ?
we calculate the entropy of the spacetime in which $\beta^2$ term
is appeared. According to the Bekenstein-Hawking entropy theorem
the black hole entropy is equal to quarter of its surface area for
which we will have
\begin{equation}S=S_0(n+\frac{1}{2})^4,~~~S_0=\pi r_0^2=\frac{\pi \beta^2
G}{M^2c^4}.\end{equation}  This shows
$S_0(\beta<0)=S_{0}(\beta>0)$ which is entropy of the system in
its ground state $n=0.$ In the next section we obtain classical
solutions of the metric field $X(r)$ and the electric potential
field $\phi(r)$ by solving the corresponding Euler-Lagrange
equations and then investigate locations of possible horizons.
\section{Classical solutions of
the fields} In the previous sections we saw that the gauge field
has not effects on the dynamics of the spacetime in the canonical
quantum gravity approach and hence we set $\gamma=-\beta$ to
remove gauge field part in the Lagrangian density (\ref{lag}). By
calculating the Euler-Lagrange equations of the fields
$\phi(\tau)$ and $X(\tau)$ in which $\tau=\ln(r/D)$ given by
(\ref{dot}), we obtain
\begin{equation}\label{41}\frac{1}{X}=\frac{4\dot{\phi}}{\phi}\end{equation} for $X$ and \begin{equation}\label{42}2\ddot{\phi}+\beta\phi\bigg[\frac{2\dot{X}}{X}-\frac{1}{X}+1
\bigg]=0\end{equation}for $\phi.$ By eliminating $\phi$ between
the above equations we obtain
\begin{equation}\label{xd}\dot{X}=\frac{1-8\beta X+8\beta X^2}{4(1-4\beta X)}\end{equation}
and
\begin{equation}\label{phid}\frac{d\phi}{\phi}=\frac{(1-4\beta X)dX}{X(1-8\beta X+8\beta X^2)}.\end{equation}
 The latter equation is obtained from $\dot{\phi}$
divided by $\dot{X}.$  The equation (\ref{xd}) reads easily to the
integral equation \begin{equation}\label{ss}\tau(X)=4\int
dX\bigg(\frac{1-4\beta X}{1-8\beta X+8\beta
X^2}\bigg)\end{equation} and (\ref{phid}) gives us
\begin{equation}\label{phs}\bigg(\frac{1}{Z}\bigg)^2-\bigg(\frac{1}{X}-4\beta\bigg)^2=8\beta(1-2\beta),~~~Z=\frac{\phi}{\phi_0}\end{equation}
 in which $\phi_0=\phi(X=1)$
is constant of integration. By substituting $X$  from the above
equation into the equation (\ref{41}) we obtain
\begin{equation}\dot{Z}=\beta Z\pm\sqrt{\frac{1}{16}+\beta\bigg(\beta-\frac{1}{2}\bigg)Z^2}
\end{equation} which reads to the following integral equation
\begin{equation}\label{tphi}\tau_\pm(\phi)=\int\frac{dZ}{\beta\pm\sqrt{\frac{1}{16}+\beta\big(\beta-\frac{1}{2}\big
)Z^2}}.\end{equation} The equations (\ref{ss}) and (\ref{tphi})
have closed form analytic solutions such that
\begin{equation}\label{tau}\tau(X)=\ln\bigg(\frac{r}{D}\bigg)=\frac{2\tanh^{-1}\bigg[\frac{(2X-1)}{
\sqrt{1-1/2\beta}}\bigg]}{\frac{1}{\sqrt{1-1/2\beta}}}\end{equation}$$-\ln[C_1(8\beta
X^2-8\beta X+1)]$$ and
\begin{equation}\tau_{\pm}(\phi)=\ln\bigg(\frac{r}{D}\bigg)=\ln[C_2(8\beta^2 z^2-1)]\pm\ln\bigg(\frac{1+2\sqrt{2\beta}z}{1-2\sqrt{2\beta}z}\bigg)\end{equation}
$$\pm\ln\bigg[\frac{2\sqrt{2\beta}(2\beta-1)z+1+\sqrt{16\beta^2(2\beta-1)z^2+2\beta}}{2\sqrt{2\beta}(2\beta-1)z-1-\sqrt{16
\beta^2(2\beta-1)z^2+2\beta}}\bigg]$$$$
\pm\sqrt{\frac{2(2\beta-1)}{\beta}}\ln\bigg[\frac{\sqrt{2\beta-1}}{\sqrt{8\beta(2\beta-1)}+\sqrt{8\beta(2\beta-1)z^2+1}}\bigg]$$
respectively in which $C_{1,2}$ are constants of integration. One
can show that the solution (\ref{tau}) can be rewritten  in terms
$r$ such that
\begin{equation}\label{511}\frac{2X-1}{\sqrt{1-\frac{1}{2\beta}}}=
\frac{[
(2X-1)^2-(1-\frac{1}{2\beta})]^\frac{1}{\sqrt{1-\frac{1}{2\beta}}}-\big(\frac{\tilde{D}}{r}\big)^\frac{1}{\sqrt{1-\frac{1}{2\beta}}}}{
[(2X-1)^2-(1-\frac{1}{2\beta})]^\frac{1}{\sqrt{1-\frac{1}{2\beta}}}+\big(\frac{\tilde{D}}{r}\big)^\frac{1}{\sqrt{1-\frac{1}{2\beta}}}}\end{equation}
where
\begin{equation}\tilde{D}=\frac{D}{2\beta C_1}.\end{equation} One can see that for $r>>\tilde{D}$ the above equation reduces to
 \begin{equation}\label{xinfty}X(\infty)=\frac{1}{2}\bigg[1+\sqrt{1-\frac{1}{2\beta}}
\bigg],~~~\beta\geq\frac{1}{2}; \beta<0\end{equation} which
reduces to asymptotically flat $X(\infty)\approx1$ for
$\pm\beta\to\infty.$ In this limits the equation (\ref{511})
reaches to the asymptotically flat metric solution
\begin{equation}\label{514}\lim_{\beta\to\pm\infty}X_{\pm}(r)\sim1\pm
\sqrt{\frac{R_{Sch}}{r}},~~~R_{Sch}=-\tilde{D}=\frac{D}{2\beta
C_1}\equiv \frac{2GM_{ADM} }{c^2}\end{equation} and unacceptable
solution $X=0.$ In fact the latter solution say about the horizon
equation of the spacetime without to say where is position of the
spacetime horizons?.
 By substituting the asymptotic metric solution (\ref{514}) into the equation (\ref{phs}) and keeping the limits $\pm\beta\to\infty$
  we obtain asymptotic solution for the
 electric potential such that \begin{equation}\label{515}\lim_{\beta\to\pm\infty}|Z(r)
 |\sim\frac{1}{2}\bigg(\frac{\pm1}{8\beta}\bigg)^\frac{1}{2}\bigg(\frac{R_{Sch}}{r
 }\bigg)^\frac{1}{4}\end{equation} which differs with the electric monopole potential.
 It may be useful to evaluate approximations about the electric charge density and total electric charge by
 calculating Poisson's equation $r^{-2}\frac{d}{dr}(r^2X(r)\frac{d Z(r)}{dr})=\rho_\beta$  which by substituting (\ref{514}) and (\ref{515})
  reads
 \begin{equation}|\rho_\beta(r)|\sim\bigg(\frac{-1}{32 R^2_{Sch}}\bigg)\bigg(\frac{\pm1}{8\beta}\bigg)^\frac{1}{2}
 \bigg(\frac{R_{Sch}}{r}\bigg)^2\bigg[5\bigg(\frac{R_{Sch}}{r}\bigg)^\frac{1}{4}\pm3\bigg(\frac{R_{Sch}}{r}\bigg)^\frac{-1}{4}\bigg]
 \end{equation} and total electric charge is \begin{equation}Q_\beta=\int_{3-space}
 d^3r\sqrt{g}\rho_\beta(r)=4\pi\int_0^{R_{Sch}}
 r^2dr\rho_\beta(r)
 \end{equation}$$=\bigg(\frac{-\pi R_{Sch}}{8}\bigg)\bigg(\frac{\pm1}{8\beta}\bigg)^\frac{1}{2}\bigg[\frac{20}{3}\pm\frac{12}{5}\bigg]$$
 where $+(-)$ sign is used for large $\beta>0(<0)$.
By substituting $R_{Sch}=\frac{2GM_{ADM}}{c^2}$ the above total
charge can be rewritten
\begin{equation}\frac{Q_\beta}{M_{ADM}}=-\frac{\pi G}{8c^2}\bigg(\frac{\pm1}{8\beta}\bigg)^\frac{1}{2}\bigg[\frac{20}{3}\pm\frac{12}{5}\bigg]
=constant.
\end{equation} This shows that for an electrostatic star the total net charge is proportional with its total mass which obeys result
of the published work \cite{Nes} in which authors proved that `charge to mass ratio` of a charged spherical static star is given by
\begin{equation}\frac{Q}{M}=\frac{2\pi\varepsilon_0G(m_p-m_e)}{e}\end{equation}
 in which $\varepsilon$ is electric permittivity of vacuum, $e$ and $m_e$ is the net charge and mass of electron, $m_p$ is the proton  mass.
 If the mass $M$ is given in solar masses and charge $Q$ in Coulombs, then it is obtained $Q=77.043M.$ Comparing these charge to mass  ratio
 for the electrostatic stars which are obtained  with different approaches a numerical value for $\beta$  parameter of the gravity model under
consideration can be predicted by the  characteristics of
electrons and protons such that \begin{equation}\beta=\frac{\pm
e^2\bigg[\frac{5}{3}\pm\frac{3}{5}\bigg]
}{512\varepsilon^2_0c^4(m_p-m_e)^2}
\end{equation} which by substituting numerical values of the electron/proton characteristics in SI units
such that $\varepsilon_0\approx8.85\times10^{-12},$
$c\approx3\times10^8,$ $m_p\approx2000 m_e$ and
$m_e\approx9\times10^{-31}$ we obtain
\begin{equation}\beta_-\approx24.533,~~~\beta_+\approx52.133.\end{equation}
By substituting these numerical values  into the equation
(\ref{xinfty}) we obtain
\begin{equation}X_-(\infty)=0.9948575235,~~~X_+(\infty)=0.9975965095\end{equation}
which has good agreement with absolutely flat Minkowski metric
$X(\infty)=1.$  Their small deviations as $\Delta
X_-(\infty)=1-X_-(\infty) \approx0.005$ and $\Delta
X_+(\infty)=1-X_{+}(\infty)\approx0.002$ can be related to
expansion of the universe or de Sitter cosmological horizon. After
providing some arguments about validity of our calculations and
comparing with work of other researchers, we bring now arguments
to say that the singular solution $X_-(r)$ is not physical but
regular solution $X_+(r)$ is. The asymptotic metric solution
(\ref{514}) describes a regular solution for electrostatic star
$X_+(r)$ and a singular solution $X_-(r)$ for a black hole with
Schwarzschild radius $R_{Sch}.$ As we saw in the previous sections
that the canonical quantum gravity approach of the model
determined that $X_+$ is stable more with respect $X_-.$ Thus we
can have an argument that full canonical quantum gravity approach
of the model under consideration prevents from having a black hole
with causal and apparent singularities. Also we saw that the
`correspondence principal` predicts in weak field limit that $X_-$
is possible to exist as physical too and so in the latter case the
causal singularities should be covered by surface of the horizons
by according to the cosmic censorship hypothesis. To show that
$X_-(r)$ is not exist as valid solution even in the classical
regimes of the fields we study here behavior of the metric field
$X(\phi)$ in the phase space without to use coordinates system. To
do so one can show that for large $\beta$ the metric solution in
phase space given by (\ref{phs}) reaches to the following
hyperbole.
\begin{equation}\bigg(\frac{1}{\bar{X}}-1\bigg)^2-\frac{1}{\bar{Z}^2}=1\end{equation} where we defined
\begin{equation}\bar{X}=4\beta X,~~~\bar{Z}=8\beta Z.
\end{equation} We plotted its diagram in figure 1 which shows there is not any point crossed with horizontal axis $\bar{Z}.$ This
means that there is not a finite electric potential for which the
horizon equation has real roots $\bar{X}=0.$  In other words there
is not horizon for the spacetime and so we must be keep $X_+(r)$
as physical asymptotic  solution and then $X_-(r)$ should
 be discarded. However if asked important question: What happens the spacetime at central
region of the minisuparspace namely $(\bar{X},\bar{Z})\to(0,0)$ at
the classical regime of the fields $\bar{Z}(\bar{X})$ which is
given by center in the figure 1, we address readers the first
section of this work, namely perfect canonical quantum gravity
behavior of the central regions of the minisuperspace $(X,\phi)$
where the spacetime behaves as simple harmonic quantum Oscillator
with stable behavior and free of every singular horizons.
\section{Concluding remarks} Here we used an alternative  generalized nonminimally coupled
Einstein-Maxwell gravity instead of the well known minimally gauge
invariant Einstein-Maxwell theory to study stability of a
spherically symmetric static curved spacetime via canonical
quantum gravity approach. Motivation of this exotic gravity model
is because of need of conformal breaking of the electromagnetic
fields which is applicable to describe origin of the cosmic
magnetic field in our expanding Universe. By calculating
Lagrangian density and Hamiltonian constraint of the system we
obtained that at asymptotically flat region of the spacetime, the
Hamiltonian constraint behaves similar to Hamiltonian of a free
particle. While at central region of the spacetime where the
curvature is not negligible, the Hamiltonian constraint of the
system behaves similar to a quantum system of simple harmonic
Oscillator. We solved Wheeler-DeWitt wave equation and obtained
energy eigenstates for a non-zero ADM mass of the system. Also we
showed that the eigenfunctionals are described by Hermit
polynomials. We also obtained quantization condition on the metric
field whose most probability reads ground state where the
curvature of the spacetime is not negligible and there is not
obtained event and apparent horizons for the spacetime. Thus one
can infer that stability of the system reaches to a stellar object
metric without the horizon which we called it `electrostatic`
stars. In fact this work confirms results of our previous work
which recently we showed stability of this metric field via
classical perspective \cite{HTF}. To check that do this
mathematical calculations are valid physically? we used Bohr`s
correspondence principal to extract semi classical metric
solutions from the quantized metric eigenstates. Also we solved
Euler Lagrange equations of the fields to obtain exact analytic
solutions of the metric field and electric potential field too.
Fortunately our quantized non singular metric solutions reach to
asymptotically flat semiclassical solutions without the horizons
and this confirms the Birkhoff theorem. To compare results of this
work with other alternatives, we obtained that the net charge of
the obtained electrostatic star is linearly depended to its ADM
mass which is approved  with results of other methods. In fact
this property is for a spherical electrostatic star and originates
form plasma behavior of interior matter of the star. Also we check
that although the model has not gauge invariance symmetry but the
gauge field has not dynamical effects on the Wheeler DeWitt wave
of the quantum system and the classical solutions of the fields
too. In this work we did not considered time dependent
perturbations for which the magnetic field is dominated same as
electric field. This is our future work which we like to
investigate. Other exotic models of Einstein Maxwell gravity is
our future aim which we like to consider still as next work to
compare with results of this work. This will done in presence and
absence of effects
of magnetic monopoles same as \cite{GHN}.\\
\vskip 0.1 cm
\textbf{Acknowledgment} \\
\vskip 0.1 cm
 I would like to thank editorial team and anonymous referees
for his/her
useful expert comments which cause to improve this work.\\
\vskip0.1 cm
\textbf{Data availability statement}\\
\vskip 0.1 cm
 All data that support the findings of this study are
included
within the article (and any supplementary files).\\
\vskip 0.1 cm
\textbf{ORCID iDs}\\
\vskip 0.1 cm Hossein Ghaffarnejad:
https://orcid.org/0000-0002-0438-6452

\begin{figure}
\centering\subfigure[{}]{\label{1011}
\includegraphics[width=0.45\textwidth]{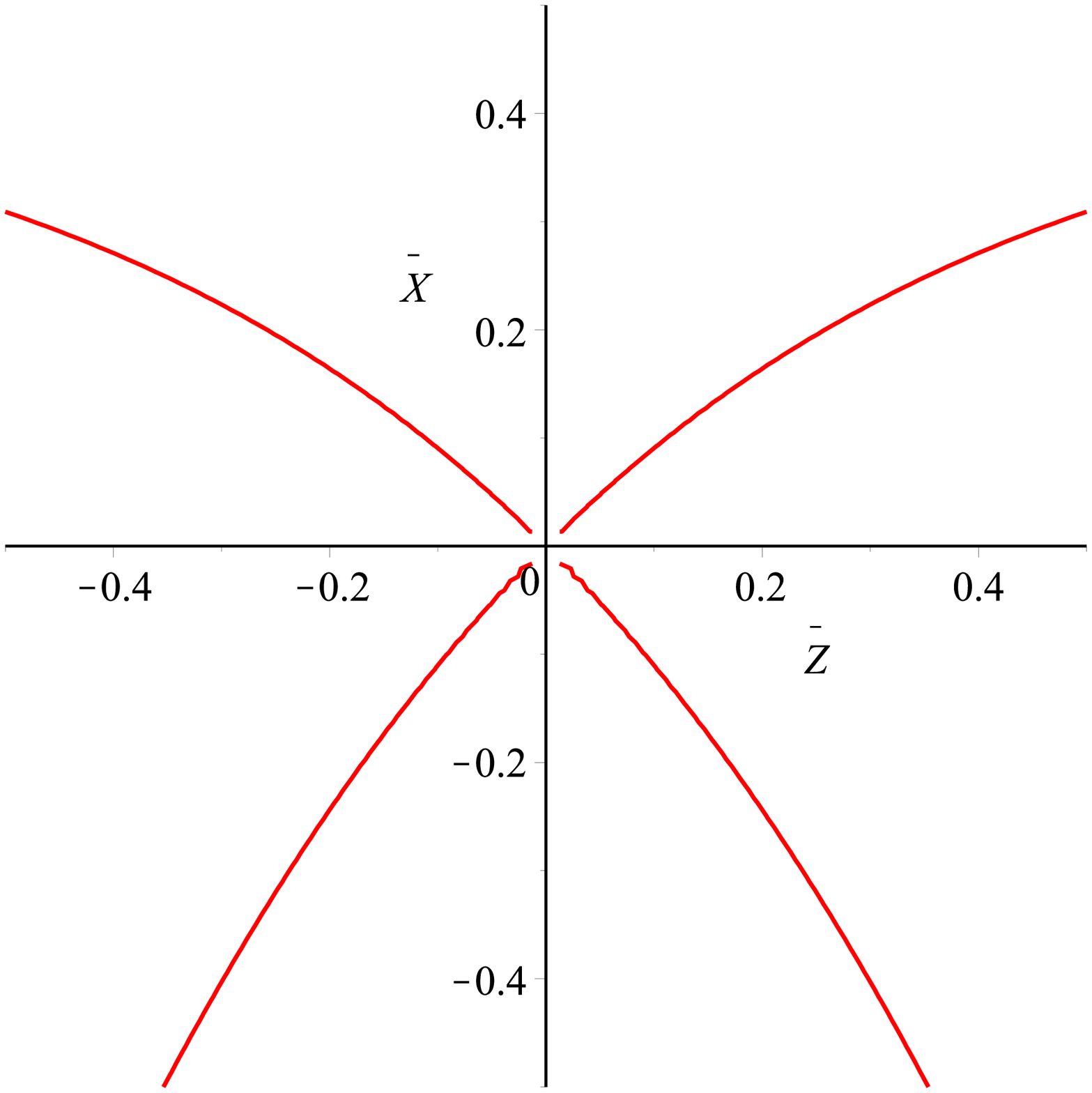}}x
\hspace{2mm} \caption{\footnotesize{Metric field values vs the
electric potential values in phase space}}
\end{figure}
\end{document}